\documentclass[aps,prd,12point,twocolumn,nofootinbib,showpacs,superscriptaddress]{revtex4-2}
\def\theequation{\arabic{section}.\arabic{equation}}
\usepackage{psfrag}
\usepackage{subfigure}
\usepackage{color}
\usepackage{mathrsfs}
\usepackage{graphicx}
\usepackage{amssymb, bm}
\usepackage{amsmath, amsthm}
\usepackage{epstopdf}
\usepackage{hyperref}
\usepackage{enumerate}
\usepackage{longtable}

\usepackage{amsfonts}
\usepackage[T1]{fontenc}
\usepackage{bm}
\usepackage{amsmath}  
\usepackage{graphicx} 
\usepackage{color}
\usepackage{appendix}

\newcommand{\be}{\begin{equation}}
\newcommand{\ee}{\end{equation}}

\begin{document}
\def\theequation{\arabic{section}.\arabic{equation}} 
%\preprint{}

\title{Asymptotic flatness and Hawking quasilocal mass}

\author{Valerio Faraoni}
\email[]{vfaraoni@ubishops.ca}
%\thanks{}
\affiliation{Department of Physics and Astronomy, Bishop's University, 
2600 College Street, Sherbrooke, Qu\'ebec, 
Canada J1M~1Z7}

\author{Andrea Giusti}
\email[]{agiusti@ubishops.ca}
%\thanks{}
\affiliation{Department of Physics and Astronomy, Bishop's University, 
2600 College Street, Sherbrooke, Qu\'ebec, 
Canada J1M~1Z7}

\author{Tyler F. Bean}
\email[]{tbean182@ubishops.ca}
%\thanks{}
\affiliation{Department of Physics and Astronomy, Bishop's University, 
2600 College Street, Sherbrooke, Qu\'ebec, 
Canada J1M~1Z7}


%\date{\today}

\begin{abstract}

We point out an association between anomalies in the Hawking quasilocal 
mass (or, in spherical symmetry, in its better known version, the 
Misner-Sharp-Hernandez mass) and unphysical properties of the spacetime 
geometry. While anomalous behaviours show up in certain quantum-corrected 
black holes, they are not unique to this context and signal serious 
physical pathologies of isolated gravitating systems in general.

\end{abstract}

% insert suggested keywords - APS authors don't need to do thisW
%\keywords{}

\maketitle

\section{Introduction}
\setcounter{equation}{0}
\label{sec:1}

According to the Equivalence Principle, which constitutes the foundation 
of general relativity (GR) and of metric theories of gravity 
\cite{Wald,Willbook}, the gravitational field can be eliminated locally 
and it is impossible to assign a local energy density to the gravitational 
field. For isolated systems, one can consider the notion of mass at 
spatial infinity, which is embodied by the Arnowitt-Deser-Misner (ADM) 
construct. This concept, however, is not defined for non-isolated systems 
(for 
example, massive objects embedded in cosmological spacetimes) and it is 
only defined asymptotically. It is, however, possible to define the 
mass-energy of a gravitating system in a quasilocal way. In the presence 
of spherical symmetry, the Misner-Sharp-Hernandez mass \cite{MSH1,MSH2} 
has been used for a long time, especially in the context of the 
gravitational collapse of fluids. The Misner-Sharp-Hernandez mass finds a 
generalization to non-spherically symmetric spacetimes in the Hawking 
quasilocal mass \cite{Hawking,Hayward}, and several other definitions of 
quasilocal energy have been proposed (see Ref.~\cite{Szabados} for a 
review).

The Hawking quasilocal mass is not normally associated with asymptotic 
flatness, however one can associate certain ``anomalies'' in the behaviour 
of the Hawking mass when the gravitational field exhibits pathologies. The 
purpose of this work is to illustrate this association and to discuss how 
the Hawking/Misner-Sharp-Hernandez mass can signal unphysical properties 
of spacetime.

The first occurrence of this association is in the context of regular 
black holes. In the quest to avoid spacetime singularities, proposals have 
been made to quantize the full GR theory or, from more phenomenological 
points of view, at least its black holes to remove the timelike 
singularities hiding inside them. Naturally, much attention has focused on 
removing the singularity of the prototypical Schwarzschild black hole and 
the quantum-corrected black holes proposed in the literature are usually 
static and spherically symmetric geometries. Often, these 
quantum-corrected black holes do not describe isolated systems {\em in 
vacuo} and, sometimes, they are not even asymptotically flat. The Bardeen 
regular black hole \cite{Bardeen} can be construed as a solution of the 
Einstein equations coupled to non-linear electrodynamics, thus it is not a 
vacuum solution \cite{AyonBeato}.  Many other examples of regular black 
holes have been provided over the years, including the more recent Planck 
star proposal (\cite{regularBH}, see 
\cite{Ansoldireview} for a review) and the subject is a mature one with a 
relatively large literature devoted to it.  Quantum-correcting the 
Schwarzschild black hole according to Loop Quantum Gravity produces a 
geometry \cite{AA1,AA2,AA3} that fails to be truly asymptotically flat 
\cite{MariamSuddho,ValerioAndreaSymmetry}. This fact causes the black hole 
geometry to exhibit unexpected unphysical properties, due to the fact that 
the small quantum gravity corrections actually dominate in regions in 
which gravity is weak, as well as in strong gravity regions near the 
singularity that they are designed to eliminate 
\cite{MariamSuddho,ValerioAndreaSymmetry}. This fact is responsible for 
unphysical properties, which include a vanishing quasilocal mass as seen  
from spatial infinity, instead of the positive Schwarzschild mass that one 
expects to recover far away from the black hole 
\cite{ValerioAndreaSymmetry}. In addition, no initially outgoing timelike 
geodesic can reach $r=+\infty$, where $r$ is the areal radius 
\cite{ValerioAndreaSymmetry}. 

Motivated by the example of quantum-corrected and regular black holes, we 
consider the more general question of whether possible variations in the 
definition of asymptotic flatness ({\em i.e.}, in the falloff rate of the 
fields) can be physically meaningful. We use the ADM mass at infinity and 
the Hawking quasilocal mass as tools to discuss physical properties of the 
gravitating systems described. The result is that the falloff rates of the 
physical fields required in the definition of asymptotic flatness are 
strictly necessary and relaxing them causes physical pathologies, which 
will be discussed.

In Sec.~\ref{sec:2} we recall the definition of ADM mass and discuss the 
physical implications of relaxing the falloff rates of the fields in it. 
Since the ADM mass is only defined at infinity, in Sec.~\ref{sec:3} we 
seek furher physical insight by using the Hawking quasilocal mass 
 \cite{Hawking, Hayward}, which is defined at any finite distance from a 
self-gravitating body, but reduces to the ADM mass at spatial infinity. It 
is also defined in non-asymptotically flat geometries, which allows us to 
explore easily geometries that relax the requirements of asymptotic 
flatness.  

We first consider spherical symmetry, in which case the Hawking  
mass reduces to the better known Misner-Sharp-Hernandez mass used in fluid 
mechanics and in gravitational collapse \cite{MSH1,MSH2}. Then, in 
Sec.~\ref{sec:4} we relax the assumption of spherical symmetry. 
Predictably, it is much more difficult to prove precise statements in this 
general situation, but we provide an argument in general ({\em i.e.}, 
non-spherically symmetric) geometries pointing again to the fact that the 
conditions in the definition of asymptotic flatness cannot be relaxed 
without introducing physical pathologies. These pathologies are reflected 
in anomalies in the Hawking mass, such as its vanishing or divergence at 
spatial infinity, or the fact that it receives a contribution from matter, 
but not from the gravitational field.

Throughout this work, we follow the notation of Ref.~\cite{Wald}. Units 
are such that the speed of light and Newton's constant are unity.

\section{Asymptotic flatness and ADM mass}
\setcounter{equation}{0}
\label{sec:2}

Let us consider the $3+1$ foliation of a general spacetime $(\mathcal{M}, 
g)$, with $g$ denoting the metric tensor, in terms of $3$-dimensional 
spacelike hypersurfaces $\Sigma _t = \{x^\mu \, | \, t (x^\mu) = 
\mbox{const.}\}$, with $t$ denoting a time function. The time evolution of 
the system is, therefore, generated by the vector field $\partial / 
\partial 
t$ that can be split into a component tangent to $\Sigma _t$ and a normal 
to the hypersurface, {\em i.e.}, 
\be 
\left( \frac{\partial}{\partial t} 
\right)^a = N \, n^a + N^a \, , 
\ee 
with $N$ the lapse function, $N^a$ the shift vector, and $n^a$ the normal 
to $\Sigma _t$ (in the coordinate representation $n_{\alpha}\sim \partial 
_\alpha t$).

The pull-back of $g$ onto $\Sigma _t$ defines the induced metric 
$\gamma_{ab} = 
\varphi ^\ast g_{ab}$, with $\varphi$ denoting the embedding of 
$\left( \Sigma 
_t, \gamma \right)$ into $\left( \mathcal{M}, g \right)$.   
~$\gamma_{ab}$, adapted to the 
coordinates on $ \left( \mathcal{M}, g \right)$, reads
\be
\gamma_{\mu \nu} = g_{\mu \nu} + n_\mu \, n_\nu \, ,
\ee
in fact $\gamma^\mu _{\,\, \nu}$ acts as a tangential projector onto $\Sigma 
_t$, {\em i.e.}, if $V^a \in T\mathcal{M}$, then $\gamma^\mu _{\,\, \nu} 
V^\nu$ 
belongs to $T\Sigma _t$. In a similar way, the Levi-Civita connection 
$\nabla$ defined on $\left( \mathcal{M}, g \right)$ induces the 
Levi-Civita connection 
$D$ on $(\Sigma _t, \gamma)$. Furthermore, if $\epsilon = \sqrt{- g} \, dx^0 
\wedge dx^1 \wedge dx^2 \wedge dx^3$ denotes the volume form on 
$\left( \mathcal{M}, g \right)$, then $\bar{\epsilon} = \sqrt{\gamma} \, 
dy^1 \wedge 
dy^2 
\wedge dy^3$. If we adapt our chart on $\Sigma _t$ so that $y^i = x^i$ for 
$i=1,2,3$, then $\bar{\epsilon} _{\alpha \beta \gamma} = n^\mu \, 
{\epsilon} _{\mu \alpha \beta \gamma}$.

Finally, the extrinsic curvature of $\left( \Sigma _t, \gamma \right)$ in 
$\left( \mathcal{M}, g \right) $ is defined as
\be
K_{ab} := \frac{1}{2} \, \mathcal{L} _n \gamma_{ab} \,.
\ee
In a coordinate chart of $\left( \mathcal{M}, g \right)$, this reads
\be
K_{\mu \nu} = \gamma_{\mu} ^{\,\, \alpha} \nabla _\alpha n_{\nu} =  
\gamma_{\mu} 
^{\,\, \alpha} \gamma_{\nu} ^{\,\, \beta} \nabla _\alpha n_{\beta} \, .
\ee

Similarly to the case of spacelike 3-surfaces $\Sigma$, one can embed 
closed 2-surfaces $\mathcal{S}$ into $\Sigma$. The normal bundle $T^\perp 
\mathcal{S}$ of $\mathcal{S}$ can be spanned by a timelike vector field 
$n^a$ and a spacelike vector field $s^a$. Usually, one also conventionally 
chooses these two vectors to be orthogonal, {\em i.e.}, $n^a s_a = 0$. 
Thus, if $\Sigma$ is a spacelike 3-surface embedded in the spacetime 
$\left( \mathcal{M}, g \right)$, one can identify $n^a$ with the 
(timelike) normal to 
$\Sigma$, whereas $s^a$ will be the normal to $\mathcal{S}$ tangent to 
$\Sigma$, {\em i.e.}, $s^a \in T \Sigma$ and $n^a \in T^\perp \Sigma$. 
(Alternatively, $T^\perp \mathcal{S}$ can be split at each $p \in 
\mathcal{S}$ in terms of two null normal vectors tangent to ingoing and 
outgoing null geodesics.)  Hence, $\left( \mathcal{S}, q \right)$ 
is an embedded closed 2-surface in $\left( \Sigma _t, \gamma \right)$, 
with $q$ being the 
pull-back of $\gamma$ to $\mathcal{S}$ that, in a coordinate chart of 
$\left( \mathcal{M}, g \right)$ adapted to the $3+1$ splitting, reads 
\be 
q_{\mu \nu} =  g_{\mu \nu} + n_\mu n_\nu - s_\mu s_\nu = \gamma_{\mu \nu} - 
s_\mu s_\nu \, . 
\ee 
The induced Levi-Civita connection on $\left( \mathcal{S}, q \right)$ is 
denoted by 
${\! {}^2 D}$, while the surface 2-form is ${{}^2 \! \epsilon} = \sqrt{q} 
\, d  z^1 \wedge d z^2$. If one considers a coordinate chart of 
$\left( \mathcal{M}, g \right)$ adapted to the $3+1$ splitting, it yields 
${{}^2 \! 
\epsilon}_{\mu \nu} = n^\alpha s^\beta \epsilon _{\alpha \beta \mu \nu} $. Then, we denote the 
deformation tensor $\Theta ^{(v)} _{ab}$ associated with the vector field 
$v^a$ normal to $\mathcal{S}$ as 
\be 
\Theta ^{(v)} _{\mu \nu} = q^{\alpha} 
_{\,\, \mu} q^{\beta} _{ \,\, \nu} \nabla _\alpha v_{\beta} \, , 
\ee 
in the usual coordinate chart of $\left( \mathcal{M}, g \right)$ adapted 
to the $3+1$ splitting. In particular, we denote by 
\be k_{ab} \equiv 
\Theta ^{(s)} _{ab} \, , 
\ee 
the extrinsic curvature of $\left( \mathcal{S}, q \right)$ inside the 
3-slice $\left( \Sigma _t, \gamma \right)$ corresponding to the spacelike 
normal $s^a$.

Let us now move on to the notion of asymptotic flatness and the $3+1$ 
decomposition using a coordinate-based approach (see \cite{Jaramillo}  for 
further details). 
Let $\Sigma$ be a 3-dimensional spacelike slice of $\left( \mathcal{M}, g 
\right)$ with 
induced metric  $\gamma_{ab}$. $\Sigma$ is an asymptotically flat slice if 
there exists a 
Riemannian background metric $f_{ij}$ such that:\\
i) $f_{ij}$ is flat, except on a 
compact domain $\mathcal{D}\subset \Sigma$;\\
 ii)  $\exists$  a Cartesian-like chart $ \{x^i \, : \, 
\mathcal{M} \to \mathbb{R}^3 \}$ such that, outside 
$\mathcal{D}$, one has $f_{ij} = \texttt{diag}(1,1,1)$ and $r \equiv 
\sqrt{x^2+y^2+z^2}$ can take arbitrary large values;\\
iii) As $r \to \infty$, one has
\begin{eqnarray}
&& \gamma_{ij} = f_{ij} + \mathcal{O} (1/r) \, ,\label{asym1}\\
&&\nonumber\\
 \quad \partial _k 
&& \gamma_{ij} = \mathcal{O} (1/r^2) \, , \\
&&\nonumber\\
&& \quad K_{ij} = \mathcal{O} (1/r^2) \, ,\\
&&\nonumber\\
&& \partial _k K_{ij} = \mathcal{O} (1/r^3) \, .\label{asym4}
\end{eqnarray}
Given an asymptotically flat spacetime foliated by asymptotically flat (or 
Euclidean) slices $\Sigma_t$, one defines spatial infinity as $r \to 
\infty$ and denotes it by $i^0$.

Let $\mathcal{V} \subset \mathcal{M}$ be a 4-dimensional spacetime region 
with boundary $\partial \mathcal{V}$ 
such that
\be
\partial \mathcal{V} = \Sigma_{t_1} \cup (-\Sigma_{t_2}) \cup \mathcal{T} 
\, ,
\ee
with $t_1 < t_2$, $\Sigma_{t_1}, \Sigma_{t_2}$ two spacelike 3-slices (as 
above) with metric and extrinsic curvature $\left(\gamma_{ab}, 
K_{ab}\right)$, 
$\mathcal{T}$ an outer timelike tube, and let the boundary condition 
be $\delta 
g_{ab} |_{\partial \mathcal{V}} = 0$. Note that $\mathcal{S}_t \equiv 
\Sigma_{t_2} \cap \mathcal{T}$ forms a closed spacelike 2-surface with 
induced metric and extrinsic curvature $\left( q_{ab}, k_{ab} \right)$.

The Einstein-Hilbert action, including also the Gibbons--Hawking--York 
boundary term, reads
\be
S = \frac{1}{16 \pi} \int _\mathcal{V} \epsilon R + \frac{1}{8 \pi} \int 
_{\partial\mathcal{V}} \bar{\epsilon} \, (K - K_0) \, ,
\ee
with $K_{0}$ denoting the extrinsic curvature of the boundary embedded in 
flat spacetime. This action then reduces to
\begin{eqnarray}
S &=& \frac{1}{16 \pi} \int _{t_1} ^{t_2} dt \left[ \int _{\Sigma_t} N 
({{}^3 \! R} + K_{ij} K^{ij} - K^2) \sqrt{\gamma} \, d^3 x 
\right. \nonumber\\
&&\nonumber\\
&\, & \left. + 2 \oint _{\mathcal{S}_t} (k - k_0) \, N \sqrt{q} \, d^2x 
\right] \, ,
\end{eqnarray}
with $k$ and $k_0$ being the trace of the extrinsic curvature of $\mathcal{S}_t$
embedded in $(\Sigma _t, \gamma)$ and $(\Sigma _t, f)$, respectively.

Moving to the Hamiltonian formalism, one finds the total Hamiltonian 
\begin{eqnarray}
H &=& - \frac{1}{16 \pi} \left\{ \int _{\Sigma_t} (N \mathcal{H} 
+ 2 N^i  \mathcal{H}_i) \, \sqrt{\gamma} \, d^3 x \right. \nonumber\\
&&\nonumber\\
&\, & \left.  + 2 \oint_{\mathcal{S}_t} [N (k - k_0) - 
N^i (K_{ij} - K \gamma_{ij}) s^j ] \, \sqrt{q} \, d^2 x \right\} 
\,,\nonumber\\
&& \label{twointegrals}
\end{eqnarray}
with $\mathcal{H} = {{}^3 \! R} - K_{ij} K^{ij} + K^2$ and $\mathcal{H}_i 
= D_j K^j _{\,\, i} - D_i K$.

{\em In vacuo}, it is $\mathcal{H} = \mathcal{H}_i = 0$ (Hamiltonian and 
momentum  constraints) on solutions of the Einstein equation.  Hence, 
on-shell, one has
\be
H_{\rm on-shell} = - \frac{1}{8 \pi} \oint _{\mathcal{S}_t} [N (k - k_0) 
- N^i (K_{ij} - K \gamma_{ij}) s^j ] \, \sqrt{q} \, d^2 x \,.
\ee
Choosing $\partial / \partial t$ so that it is associated with some 
asymptotically inertial observer, {\em i.e.}, $N=1$ and $N^i = 0$ when $r 
\to \infty$, yields the ADM mass
\be
M = - \frac{1}{8 \pi} \lim _{\mathcal{S}_t (r\to \infty)} \oint 
_{\mathcal{S}_t} (k - k_0)  \, \sqrt{q} \, d^2 x \, ,
\ee
and then using the asymptotically flat slicing one finds
\be
M = \frac{1}{16 \pi} \lim _{\mathcal{S}_t (r\to \infty)} 
\oint _{\mathcal{S}_t} \left(\partial_j  \gamma_i ^{\,\, j} - \partial _i 
\gamma^j  
_{\,\, j} \right) s^i \, \sqrt{q} \, d^2 x 
\ee
(see Ref.~\cite{HawkingHorowitz} for an explicit derivation).
The asymptotic flatness conditions guarantee the convergence of 
this integral.

To appreciate the effect of metric components decaying slower than $1/r$, 
it is useful 
to contemplate the analogous situation in Newtonian gravity. {\em In 
vacuo}, the 
Newtonian potential $\phi$ solves the Laplace equation $\nabla^2 \phi=0$ 
and can be expressed as the sum of a monopole term, a dipole term, {\em 
etc.}, which makes the first integral in Eq.~(\ref{twointegrals}) 
converge. The fact that $\phi$ decays slower than $1/r$ signals the 
presence of matter (or, possibly, effective matter\footnote{This is the 
case, for example, if a cosmological constant is introduced into the 
Laplace equation.}) in space, in which case the Laplace equation turns 
into the Poisson equation $\nabla^2 \phi= 4\pi \rho$. A similar property 
holds in GR: {\em in vacuo} and for a stationary self-gravitating and 
isolated source, the 
general metric is necessarily given by a multipole expansion with the 
first term scaling as $1/r$ and no terms scaling as $r^{-(1-\epsilon)}$ 
(with $\epsilon >0$) are possible \cite{Thornereview}. The curvature 
tensor 
coincides with the Weyl tensor ${ C^a}_{bcd}$, which exhibits the peeling 
property along null geodesics \cite{Penrose,Geroch}.  The failure to 
satisfy this property for a stationary 
spacetime signals the presence of matter (or effective matter) and a 
nonvanishing Ricci tensor $R_{ab}$ (see Sec.~\ref{sec:4}).

In the presence of matter fields, ${\cal H} \propto \rho$ and ${\cal H}^i 
\propto J^i$ (where $\rho $ and $J^i$ are the energy density and energy 
current density, respectively) and the first integral in the 
right hand side 
of Eq.~(\ref{twointegrals}) converges only if the matter fields decay 
sufficiently fast.  This is the case, for example, for exact 
solutions of the Einstein equations describing relativistic stars with 
energy density that is not a function with compact support but decays very 
fast as 
$r\rightarrow \infty$ (see \cite{Lakereview} for a review).  If this 
integral diverges, there cannot be asymptotic flatness and 
the ADM mass is not defined. What is more, any pathologies in the energy 
density or effective density (for example, a negative sign, as in certain 
quantum-corrected black holes) will leave an imprint in the ADM mass (when  
the latter is well-defined).

\section{Quasilocal mass---spherical symmetry}
\setcounter{equation}{0}
\label{sec:3}

Let us turn now to a different concept of mass, the 
Hawking quasilocal mass, which has the potential to provide 
extra information with respect to the ADM mass. In fact, the quasilocal 
mass is defined using topological 2-spheres of finite size, while the ADM 
mass is necessarily defined only at spatial infinity. For 
simplicity, we restrict to spherically symmetric and static geometries 
$g_{ab}$. The line element can be written as 
\be
ds^2 = -A(r)dt^2 +B(r)dr^2 +r^2 d\Omega_{(2)}^2 \label{gauge} 
\ee 
without loss of generality, where $r$ is the areal radius defined by the 
2-spheres of symmetry and $d\Omega_{(2)}^2 \equiv d\vartheta^2 +\sin^2 
\vartheta \, d\varphi^2$ is the line element on the unit 2-sphere. 

In spherical symmetry, the Hawking quasilocal mass \cite{Hawking, 
Hayward} reduces \cite{Haywardspherical} to the better known 
Misner-Sharp-Hernandez mass $M_\text{MSH}$ defined by \cite{MSH1, MSH2} 
\be
M_\text{MSH}=\frac{r}{2} \left( 1-\nabla^c r \nabla_c r \right) 
\label{MSH}
\ee
which, in the gauge~(\ref{gauge}), assumes the form
\be
M_\text{MSH}= \frac{r}{2} \left(1- \frac{1}{B} \right) \,.\label{MSHgauge}
\ee

The Loop Quantum Gravity black hole of \cite{AA1,AA2,AA3} fails to be 
asymptotically flat \cite{MariamSuddho,ValerioAndreaSymmetry} and this 
feature is reflected in a  vanishing quasilocal mass at large (areal) 
radii  \cite{ValerioAndreaSymmetry}.  Other quantum-corrected black holes 
have the correct asymptotic flatness. 
For example, the Kehagias-Sfetsos geometry is a solution of 
Ho\v{r}ava-Lifschitz gravity \cite{Horava} in the presence of plasma, with 
line element \cite{KehagiasSfetsos}
\be
ds^2 =-f(r)dt^2+\frac{dr^2}{f(r)}+r^2 d\Omega_{(2)}^2 \,, \label{sss}
\ee
where
\be
f(r)=1+ \omega_{KS} r^2 \left[ 1-\left( 1+\frac{4m}{\omega_{KS}\, r^3} 
\right)^{1/2} \right] \,.
\ee
By expanding for $m/r \ll 1$, one obtains $f(r)\simeq 1-2m/r+ 
\mathcal{O}\left( 1/r^2\right) $, which is the correct asymptotics for 
asymptotic flatness.

Let us discuss the relation between Misner-Sharp-Hernandez mass and 
asymptotic flatness more in general. In asymptotically flat spacetimes, 
the metric component $g_{rr}$ has the asymptotics
\be
g_{rr}=1+ \mathcal{O}\left(\frac{1}{r} \right)\,,
\ee
which implies that also $g^{rr}=1+\mathcal{O}(1/r)$; then  the 
quasilocal mass~(\ref{MSHgauge}) is finite since the prefactor $r$ 
cancels the only remaining term in the round brackets, which is of order 
$1/r$. 
This situation is physical and occurs, for example,  in the Schwarzschild 
geometry
\be
ds^2=- \left(1-\frac{2m}{r} \right)dt^2 +\frac{dr^2}{ 1- 2m/r } +r^2 
d\Omega_{(2)}^2\,,
\ee 
for which $M_\text{MSH} $ does not depend on the position $r$ and 
coincides with the Schwarzschild mass $m$ everywhere outside the horizon 
$r=2m$, and with the ADM and the Newtonian mass as $r\rightarrow +\infty$.

If the metric is not  asymptotically flat, say 
\be
g_{rr}=1+\mathcal{O}\left(\frac{1}{r^{1+\epsilon} }\right) 
\ee
with $\epsilon>0$, then $M_\text{MSH}(r) \rightarrow 0$ as $r\rightarrow 
\infty$.  This is the situation, {\em e.g.},  for the quantum-corrected 
Schwarzschild black hole of \cite{AA1,AA2,AA3}, for which $g_{rr}= 
\left[ 1-\left( 2m/r\right)^{1+\epsilon}\right]^{-1}$ where $\epsilon$ is 
a small positive 
number (dependent on the black hole mass) which, for a solar mass black 
hole, assumes the value $ \sim 10^{-26}$ \cite{AA3}. In this case the 
mass $M_\text{MSH}$ (which is always defined in spherical symmetry) 
vanishes as 
$r\rightarrow +\infty$. In this limit,  the Newtonian potential $\phi_N $ 
is given by
\be
1+2\phi_N = 1-\left( \frac{2m}{r}\right)^{1+\epsilon} \equiv 
1-\frac{2M(r)}{r} \,,
\ee
and one obtains the position-dependent Newtonian mass 
\be
M(r)=\left( \frac{2m}{r}\right)^{\epsilon}\, m \,,
\ee
which does not coincide with the mass obtained from the monopole term 
of the expansion of the metric in multipoles, as 
it should.

If instead $g_{rr}=1+ \mathcal{O}\left( 1/ r^{1-\epsilon} \right)$ (again, 
with $\epsilon>0$), then the quasilocal mass $M_\text{MSH}(r)$ is again 
position-dependent and diverges as $r\rightarrow +\infty$, another 
unphysical situation for an isolated object.

What is more, if the asymptotics required by the definition of asymptotic 
flatness is not satisfied, the Newtonian limit is jeopardized. In an 
asymptotically flat system, at large spatial distances from the source of 
gravity one ought to recover the post-Newtonian approximation 
\cite{Willbook} in which the line element reduces to 
\be
ds^2 =-\left( 1+\phi_N\right) dt^2 +\left( 1-\phi_N\right) \left( dr^2 
+r^2 d\Omega_{(2)}^2 \right) \,.
\ee
The dominant term in the Newtonian potential $\phi_N$ must be a monopole, 
and this term must be present. Contrary to electrostatics, in which 
electric charge can have positive or negative sign and one could have a 
dipole with zero total charge, mass cannot be negative and the first term 
in a multipole expansion of $\phi_N$ must necessarily be the monopole term 
scaling as $1/r$. The failure to obtain such a term means that the 
geometry does not admit a Newtonian limit. While this possibility is fine 
for, {\em e.g.}, gravitational waves that do not have a counterpart in 
Newtonian gravity, it is unacceptable for an isolated black hole.

Another example is given by a Reissner-Nordstrom naked singularity with 
electric charge and vanishing mass parameter,
\be 
ds^2=-\left(1+\frac{Q^2}{r^2} \right) dt^2 +\frac{dr^2}{1+Q^2/r^2} +r^2 
d\Omega_{(2)}^2 \,,
\ee
which has Misner-Sharp-Hernandez quasilocal mass
\be
M_\text{MSH}(r)=-\frac{Q^2}{2r} \,.
\ee
A silly object like an electric charge without mass violates the 
positivity of the quasilocal energy everywhere and shouldn't exist. 
Although, superficially, the metric reduces 
to the Minkowski one away from the central object, it does so with the 
wrong asymptotics $g_{rr}=1+ \mathcal{O}\left(1/r^2 \right)$, which  
creates a negative Misner-Sharp-Hernandez mass everywhere. Although 
$M_\text{MSH}$ 
is defined 
independent of the energy conditions, a deviation from the correct 
asymptotics signals the presence of a distribution of 
mass-energy incompatible 
with an isolated object and true asymptotic flatness, or some 
physical pathology. If the energy 
density and stresses of the latter do not fall off sufficiently rapidly, 
then the notion of asymptotic flatness as referring to isolated energy 
distributions fails.  What is more, if this energy distribution 
corresponds to negative energies, it leaves an imprint on the quasilocal 
mass and may make it negative.  Of course, this is not the only 
way to violate the positivity of the MSH mass: for example, the 
Schwarzschild  solution with negative mass (another naked singularity) 
does that, but it has the 
correct asymptotics required by asymptotic flatness.

The spherically symmetric Bardeen regular black hole \cite{Bardeen} is 
asymptotically flat and the MSH mass is well behaved, and so are the 
Hayward regular black hole \cite{HaywardBH} and its modification 
describing a Planck star \cite{CarloSimone}, the Peltola-Kunstatter 
black hole arising in polymer quantization of the Schwarzschild geometry 
\cite{Gabor}, and the Gambini-Olmedo-Pullin regular black hole \cite{GOP}. 
Therefore, quantum corrections do not necessarily spoil 
asymptotic flatness or introduce physical pathologies or mass anomalies.

	We can add some insight by recasting the spherical line element in 
a particular gauge exhibiting explicitly the Misner-Sharp-Hernandez mass. 
Any spherically symmetric metric can be rewritten in the Abreu-Visser 
gauge 
\be
ds^2 =-\mbox{e}^{-2\Phi} \left( 1-\frac{2M}{r}\right) dt^2 
+\frac{dr^2}{1-2M/r} +r^2 d\Omega_{(2)}^2 \,,
\ee
where $\Phi=\Phi(t,r), M=M(t,r)$ and, {\em a posteriori}, $M$ is shown to 
be the Misner-Sharp-Hernandez mass  \cite{AbreuVisser}. It follows 
immediately from this line element that, as $r\rightarrow +\infty$, 
the asymptotic flatness conditions~(\ref{asym1})-(\ref{asym4}) require 
that\footnote{This conclusion agrees with the recent 
Ref.~\cite{GudapatiYau}.}  
$2M/r=\mathcal{O}(1/r)$ and $ M$ tends to a finite limit  
$M_{\infty}$, or $M=\mathcal{O}(1)$. 

Let us consider now the stress energy tensor $T_{ab}$ associated with this 
geometry, which is given by
\begin{eqnarray}
G_{00}&=&8\pi T_{00} = \frac{2M'}{r^2} \,,\\
&&\nonumber\\
G_{01} &=& \frac{2\dot{M}\, \mbox{e}^{\Phi}}{r^2 
\left(1-2M/r\right)}  \,,\\
&&\nonumber\\
G_{11} &=& -\frac{2 M'}{r^2} -\frac{2\Phi'}{r} 
\left(1-\frac{2M}{r} \right) 
\,,\\
&&\nonumber\\
G_{22}&=&G_{33}  =\frac{M''}{r} -\frac{ \mbox{e}^{-\Phi}}{r} 
\frac{\partial}{\partial t} \left[ \frac{\dot{M} \, 
\mbox{e}^{\Phi}}{ \left(1-2M/r\right)^2 }\right] \nonumber\\
&&\nonumber\\
&\, & -\frac{ \mbox{e}^{\Phi} }{r\sqrt{1-2M/r}} 
\frac{\partial}{\partial r} \left[ r \left(1-\frac{2M}{r} \right)^{3/2} \, 
\mbox{e}^{-\Phi} \, \Phi' \right] \,\nonumber\\
&&
\end{eqnarray} 
where a prime and an overdot denote differentiation with respect to radius 
and time, respectively. Although these expressions are too cumbersome to 
draw general conclusions, 
we can restrict to static ($\dot{M}=0$) geometries for which $\Phi \equiv 
0$. Almost all the quantum-corrected black  holes proposed in the 
literature (but not Planck stars \cite{CarloSimone}) have this form. 
Then, the energy density of matter is simply
\be
\rho = \frac{M'}{4\pi r^2} 
\ee
and we conclude immediately that vacuum corresponds to constant $M$ (as in 
the case of the Schwarzschild black hole) and, in the presence of matter, 
$ \rho>0 $ if and only if the Misner-Sharp-Hernandez mass increases with 
radius, $M'>0$. Furthermore, the fact that $M$ decreases with $r$, 
{\em i.e.}, $M'<0$, 
signals the presence of a negative energy density, which decreases the 
value due to a central object that would be constant in the absence of 
this energy distribution in its exterior (this is exactly the case of the 
quantum-corrected Schwarzschild black hole of \cite{AA1,AA2,AA3}). 
Therefore, pathologies in the behaviour of the Misner-Sharp-Hernandez 
mass signal physically pathological behaviour of the geometry.

\section{Quasilocal mass---general spacetimes}
\setcounter{equation}{0}
\label{sec:4}

Let us remove now the assumption that the spacetime is spherically 
symmetric or stationary. The Misner-Sharp-Hernandez mass is then 
generalized by the Hawking quasilocal mass \cite{Hawking, Hayward}, 
defined as follows.
 
Let $S$ be a spacelike, compact, and orientable 2-surface; denote with 
$\mathcal{R}$ the induced Ricci scalar on $S$, and let $\theta_{(\pm)}$ 
and $\sigma_{ab}^{(\pm)}$ be the expansions and shear tensors of a pair of 
null geodesic congruences (outgoing and ingoing from the surface $S$). Let 
$h_{ab}$ be the 2-metric induced on $S$ by $g_{ab}$, let 
	%$\omega^a$ be the projection onto $S$ of the commutator of the 
	%null  normal vectors to $S$ \cite{Hayward},  
$\mu$ be the volume 2-form on the surface $S$, while ${\cal A}$ is the 
area 
of 
$S$; then \cite{Hawking} 
\be 
M_\text{H} \equiv \frac{1}{8\pi}\sqrt{\frac{{\cal A}}{16\pi}}\int_S 
\mu\left(\mathcal{R}+\theta_{(+)} \theta_{(-)} 
-\frac{1}{2} \, \sigma_{ab}^{(+)} \sigma^{ab}_{(-)}   
	%-2\omega_a\omega^a
\right) \,.\label{E:HH} 
\ee
As  a consequence of the Riemann 
tensor splitting into Ricci and Weyl parts \cite{Wald}
\begin{equation}
R_{abcd}=C_{abcd} + g_{a[c}R_{d]b} -g_{b[c} R_{d]a} 
-\frac{R}{3} \, g_{a[c} g_{d]b} 
\end{equation}
(where $R_{ab}$ and $C_{abcd}$ are the Ricci and Weyl 
tensors, respectively, and $R\equiv {R^c}_c$ is the Ricci scalar), the 
Hawking mass splits into two contributions, one coming from matter 
and one from 
the vacuum gravitational field, respectively. We recall this 
decomposition, performed in Ref.~\cite{Symmetry2015}. We use the 
contracted Gauss equation 
\cite{Hayward}
\begin{equation}
{\cal R}^{(h)} +\theta_{(+)} \theta_{(-)} -\frac{1}{2} \, 
\sigma_{ab}^{(+)} \sigma^{ab}_{(-)}  = h^{ac}h^{bd} 
R_{abcd} \,\label{Gauss}
\end{equation}
to compute the integral defining the 
Hawking mass.  Using then the Einstein equations 
\be
R_{ab} =8\pi G \left( T_{ab}
-\frac{1}{2} \, g_{ab}T \right) 
\ee
and $R=-8\pi G T$ (where $T \equiv {T^c}_c$), one obtains 
\begin{eqnarray}
h^{ac} h^{bd} R_{abcd} &=& h^{ac} h^{bd} C_{abcd}  
 +8\pi G h^{ac} h^{bd} \Big[ g_{a[c}T_{d]b} \nonumber\\
&&\nonumber\\
&\, & 
-g_{b[c}T_{d]a} -\frac{T}{2}  \left( g_{a[c}g_{d]b} 
-g_{b[c}g_{d]a} \right)\Big] \,. \nonumber\\
&& \label{questa}
\end{eqnarray}
Then, 
\begin{eqnarray}
&& h^{ac} h^{bd} \left( g_{a[c}g_{d]b} -g_{b[c}g_{d]a} 
\right) =2 \,,\\
&&\nonumber\\ 
&& h^{ac} h^{bd} \left( g_{a[c}T_{d]b} -g_{b[c}T_{d]a}  
\right) = h^{ab}T_{ab} \,  
\end{eqnarray}
give the Hawking mass as\footnote{This splitting, and the 
corresponding equation~(\ref{split1}), occur also in  
scalar-tensor gravity \cite{stquasilocal}.} \cite{Symmetry2015}
\begin{eqnarray}
M_\text{H} &=&  
\sqrt{ \frac{A}{16\pi}} \int_{{\cal S}}\mu
\left( h^{ab}T_{ab}- \frac{2T}{3}  \right)  \nonumber\\
&&\nonumber\\
&\, & 
+\frac{1}{8\pi G} \sqrt{ 
\frac{A}{16\pi}} \int_{{\cal S}}\mu \,
h^{ac} h^{bd} C_{abcd} \,, \label{split1}
\end{eqnarray}
where the first integral on the right hand side is the matter contribution 
and the second integral is the Weyl free field contribution, and the 
only one present {\em in vacuo}. Since we have used the Einstein 
equations, the 
rest of this discussion applies only to geometries that solve these 
equations. 

If the matter content of spacetime consists of a single perfect fluid 
with stress-energy tensor   
\be
T_{ab}=\left( P+\rho \right) u_a u_b +P g_{ab} \,, 
\ee  
energy density $\rho$, pressure $ P$, and 4-velocity $u^c$, then one can 
choose the 2-surface ${\cal S}$ comoving with the fluid ({\em i.e.}, the 
unit normal $n^a$ to ${\cal S}$ pointing outside of $\Sigma_t$ is parallel 
to the timelike fluid 4-velocity $u^a$), $h_{ac}u^c$ vanishes, and 
\begin{equation}
 h^{ab}T_{ab}- \frac{2 T}{3} = 
\frac{2 \rho}{3} 
\end{equation}

In the case of an imperfect fluid, the stress-energy 
tensor is instead
\be 
T_{ab}= \rho u_a u_b +P \gamma_{ab} +q_a u_b +q_b u_a +\Pi_{ab} \,,
\ee 
where $\gamma_{ab}$ is the 3-metric on the 
3-space orthogonal to $u^a$, as in 
\be
g_{ab}=-u_a  u_b + \gamma_{ab} \,,
\ee
$q^a$ is a purely spatial heat current 
vector ($q^c u_c=0$), and $\Pi_{ab}$ is the symmetric, trace-free, shear 
tensor. The trace is $T=-\rho +3P$ and now \cite{Symmetry2015}
\be 
h^{ab} T_{ab} -\frac{2T}{3}= \frac{2}{3}\, \rho 
+h^{ab}\Pi_{ab}= \frac{2}{3}\, \rho +{\Pi^2}_2 +{\Pi^3}_3 = 
\frac{2}{3}\, \rho -{\Pi^1}_1  
\ee 
(where $ \left( x^2, x^3 \right)$ are coordinates on ${\cal S}$).

Let us consider vacuum, in which case $M_\text{H}$ given by 
Eq.~(\ref{split1}) coincides with the Weyl contribution. In asymptotically 
flat spacetimes according to the definition of Sec.~\ref{sec:2}, the Weyl 
tensor enjoys the well-known peeling property \cite{Penrose,Geroch}. Let 
$\gamma$ denote null geodesics going from a  finite point to null 
infinity, $\lambda$ be an affine parameter along such a geodesic, and $k^a 
$ its 4-tangent. Then, the Weyl tensor splits according to 
\begin{eqnarray}
{C^{a}}_{bcd} &=& \frac{({C^{a}}_{bcd})^{(I)} }{\lambda} + 
\frac{{(C^{a}}_{bcd})^{(II)} }{\lambda^2} + 
\frac{{(C^{a}}_{bcd})^{(III)} }{\lambda^3}\nonumber\\
&&\nonumber\\
&\, &  + 
\frac{{(C^{a}}_{bcd})^{(IV)} }{\lambda^4}+ 
 \mathcal{O}\left( \frac{1}{\lambda^5}\right) \label{peeling}
\end{eqnarray}
where, in the algebraic classification of Ref.~\cite{Wald},  
$({C^{a}}_{bcd})^{(I)} $ is of type ~IV, $({C^{a}}_{bcd})^{(II)}$ of 
type~III, $({C^{a}}_{bcd})^{(III)}$ of type~II or II-II, and $k^a$ is the 
repeated principal null vector. $({C^{a}}_{bcd})^{(IV)} $ is of type~I and 
$k^a$ is one of the principal null directions of ${C^{a}}_{bcd}$.

This asymptotics in terms of an affine null geodesic parameter may not 
seem illuminating in general, but there is a situation in which it is, and 
which includes most of the regular black holes proposed in the 
literature. Let the spacetime be stationary and spherically symmetric, 
with the extra requirement that $ g_{tt} \, g_{rr}=-1 $; that is, the line 
element assumes the form~(\ref{sss}). As shown in Ref.~\cite{Jacobson}, 
this extra requirement is equivalent to the areal radius $r$ being an 
affine parameter along radial null geodesics. Now consider the surface 
${\cal S}$ to be a 2-sphere orbit of the spherical symmetry, and $\gamma$ 
to be radial outgoing null geodesics emanating from ${\cal S}$. Then, the 
peeling property~(\ref{peeling}) of the Weyl tensor can be rewritten using 
$r$ instead of $\lambda$. This equation then shows that no terms 
decreasing slower than $1/r$ are possible in the integrand of 
$M_\text{H}$ {\em in vacuo}. Such terms may be created when a form of 
matter 
(or effective matter) with $T_{ab} \neq 0$, responsible for the first 
integral in the right hand side of Eq.~(\ref{twointegrals}), produces a 
nonvanishing Ricci tensor $R_{ab}$. Similarly, no fractional powers of 
$1/r$ are possible in the Weyl tensor {\em in vacuo}.

In general ({\em i.e.}, non-spherically symmetric) geometries, the affine 
parameter $\lambda$ along null geodesics does not coincide with the radial 
coordinate (assuming that polar coordinates are used). However, in 
asymptotically flat spacetimes, the dominant term as 
$r\rightarrow \infty $ is the monopole one \cite{Thornereview} and the 
property $ 
g_{tt} \, g_{rr}=-1$ is satisfied with better and better accuracy further 
and further away from the source. Since the metric components 
$g_{\vartheta\vartheta}, g_{\varphi\varphi} $  in polar 
coordinates scale as $r^2$ and $ r^2 \sin^2 \vartheta $, respectively, we 
have
\be
 C_{2323} \sim r^2  {C^2}_{323} \sim r ( {C^2}_{323})^{(I)}   
\ee
and 
\be
h^{ac}h^{bd} C_{abcd} \simeq  \frac{2 C_{2323}}{r^4 \sin^2 \vartheta} \sim 
\frac{ ({C^2}_{323})^{(I)} }{r^3} \,.
\ee
Then, {\em in vacuo}, 
\begin{eqnarray}
M_\text{H} &=& \frac{1}{8\pi} \sqrt{ \frac{A}{16\pi}} \int_{ {\cal S}} \mu 
\, h^{ac}h^{bd} C_{abcd} \simeq  
\frac{r}{16\pi} \frac{ ({C^2}_{323})^{(I)} }{r} \nonumber\\
&&\nonumber\\
& \sim & \frac{ 
({C^2}_{323})^{(I)} }{16\pi }  \,.
\end{eqnarray}
If the system is not asymptotically flat, there will be the contribution 
to $M_\text{H}$ from the matter stress-energy tensor $T_{ab}$ and the 
peeling property of the Weyl tensor will not be satisfied. Then, the 
dominant term will not be of order $\mathcal{O}(1/r)$ and the Weyl 
contribution to $M_\text{H}$ will diverge or vanish. The latter situation 
corresponds 
to zero contribution to $M_\text{H}$ from the gravitational field, with 
$M_\text{H}$ reducing solely to the matter contribution. Both cases are 
unphysical.

\section{Conclusions}
\setcounter{equation}{0}
\label{sec:5}

Physical anomalies in the general-relativistic gravitational field can be 
signalled by anomalies of the Hawking quasilocal mass $M_\text{H}$ 
\cite{Hawking,Hayward} or, in spherical symmetry, of its better known 
version, the Misner-Sharp-Hernandez mass \cite{MSH1,MSH2}. These anomalies 
include situations in which the quasilocal mass becomes negative, zero, or 
diverges. While this association is brought about by certain 
quantum-corrected black holes, the association between anomalies in 
$M_\text{H}$ and physical pathologies is more general, as shown by the 
examples discussed in this work. In particular, a monopole term scaling as 
$1/r$ is a necessity for isolated gravitating systems and for their 
Newtonian counterparts (GR solutions which do not have Newtonian 
counterparts, or non-asymptotically flat analytical solutions that are not 
realized in nature, such as infinitely long cylindrical solutions, or 
$pp$-waves, escape this requirement).

\begin{acknowledgments} 
We thank a referee for useful comments. This work is supported, in 
part, by the Natural 
Sciences \& Engineering Research Council of Canada (Grant no. 2016-03803 
to V.F.) and by Bishop's University. The work of A.G. has been carried out in the framework of 
the activities of the Italian National Group for Mathematical Physics [Gruppo Nazionale per la Fisica
Matematica (GNFM), Istituto Nazionale di Alta Matematica (INdAM)].
\end{acknowledgments}

%\appendix
%\section{ }
%\renewcommand{\theequation}{A.\arabic{equation}}

\end{document}